\documentstyle[prl,aps, multicol,epsfig]{revtex}

\newcommand{\beq}{\begin{equation}}
\newcommand{\eeq}{\end{equation}}
\newcommand{\beqa}{\begin{eqnarray}}
\newcommand{\eeqa}{\end{eqnarray}}
\newcommand{\la}{\lambda}

\newcommand{\rh}{\rho}

\newcommand{\da}{\dagger}

\newcommand{\al}{\alpha}
\newcommand{\si}{\sigma}

\newcommand{\om}{\omega}

\newcommand{\LJ}{\langle}					    
\newcommand{\RJ}{\rangle}

\def\intt{\int^t_0}

\def\e{{\rm e}}

\begin{document}
\draft 
\title{Post-Markov master equation for the dynamics of open quantum systems}
\author{Ting Yu$^{1}$, Lajos Di\'osi$^{2,3}$, Nicolas Gisin$^1$,
and Walter T. Strunz$^4$}
   
        \address{ \protect\small\em $^1$Group of Applied Physics,
University of Geneva, 
        1211 Geneva 4, Switzerland\\
        \protect\small\em $^2$Institute for Advanced Study, Wallotstrasse 19, 
        D-14193 Berlin, Germany\\ 
        \protect\small\em $^3$ Research Institute for Particle and Nuclear
        Physics, 1525 Budapest 114, POB 49, Hungary\\
       \protect\small\em $^4$Fachbereich Physik, Universit\"at GH Essen, 
        45117 Essen, Germany}
        \date{May 3, 1999}

\maketitle
   \begin{abstract}
A systematic first-order correction to the standard Markov master equation
for open 
quantum systems interacting with a bosonic bath is presented. It
extends the Markov Lindblad  
master equation to the more general case of non-Markovian evolution. The
meaning and 
applications of our `post'-Markov master equation are illustrated with
several 
examples, including a damped two-level atom, the spin-boson model and 
the quantum Brownian 
motion model. Limitations of the Markov approximation,
the problem of positivity violation and initial slips
are also discussed.
   \end{abstract}
	
  \begin{center}
 03.65.Bz Foundations, theory of measurement, miscellaneous theories \\
 05.40.-a Fluctuation phenomena, random processes, noise, and Brownian motion \\
  42.50.Lc Quantum fluctuations, quantum noise, and quantum jumps\\
\end{center}
	\widetext
The fundamental approach to {\it open} quantum systems relies on a 
closed system-plus-reservoir model whose time evolution is 
governed by the standard Schr\"odinger or von-Neumann equation.
Due to the coupled dynamics of system and
environment, effectively, the evolution of the system's
reduced density operator $\rho_t$ depends on 
its past, it is {\it non-Markovian}.
For a quantum particle coupled to
an environment of harmonic oscillators, this is obvious from
its path integral propagator as derived by
Feynman and Vernon \cite{FeynmanVernon}.
It is also apparent from Zwanzig's projection approach \cite{Zwanzig} 
where the evolution equation for $\rh_t$ takes the form of an 
integro-differential equation involving
an integral over the whole history of the density
operator. Exactly how far back the integration over the past
has to be extended defines the
environmental `memory' time scale $\tau$.

A finite memory time $\tau$ leads to 
severe problems for the treatment of
open system dynamics which is even more true for 
attempts to find numerical solutions.
Growing experimental advances on mesoscopic scales~\cite{BLM}, however,
demand an efficient theory for open system dynamics beyond the
standard Markov approximation:
these are experiments with high-Q 
microwave cavities, investigations of the spontaneous emission from
atoms in a structured radiation continuum 
(e.g. `photonic band gap' materials) or the output coupling from a
Bose-Einstein condensate to create an atom laser,
to name a few. Also, the important phenomenon of decoherence which takes
place on time scales that can be of the same order as the correlation time 
of the environment requires theories beyond the standard Markov
approximation. 

There are a few derivations of exact non-Markovian master equations 
for model systems (see e.g. \cite{Haake,HPZ,HY_RP,AH}).
Remarkably, despite the underlying non-Markovian 
dynamics, these exact evolution equations
may be cast into the form of a {\it time-local} master equation
involving the reduced 
density operator $\rho_t$ at time $t$ only. A finite
memory time and thus non-Markovian effects 
are exactly taken into account by suitable time dependent coefficients 
entering the resulting master equation. 

The derivation of a useful
master equation for general open systems 
beyond the Markov regime remains an outstanding open problem.  
In this Letter we present a first order `post'-Markov
master equation for an arbitrary quantum
system coupled to an environment of harmonic oscillators. As in the
case of the above-mentioned soluble models, our result takes the form 
of a {\it time-local} master equation with time dependent coefficients involving
the bath correlation function.
To zeroth order in the memory time ($\tau = 0$) we recover the
standard Markov master equation.

To begin with, let us recall the Markov case. From an axiomatic
approach, Lindblad~\cite{Lindb} derived the 
most general Markov evolution equation for a density operator
($\hbar=1$):
\beq
\dot\rh_t = -i [H,\rh_t] + \frac{1}{2}\left(
[L, \rh_t L^\da] + [L\rh_t, L^\da]\right),
\label{lindb}
\eeq  
where, for simplicity, we only consider the case of a single Lindblad 
operator $L$
representing the influence of the environment. Such Markov Lindblad
master equations are widely used in quantum optics~\cite{Gar_Car}. 
We identify two `system' time scales in (\ref{lindb}): 
the ordinary dynamic time scale determined by the Hamiltonian $H$
(which we call $\omega^{-1}$), and a
damping or relaxation time scale (which we call $\gamma^{-1}$) 
determined by the operator $L^\dagger L$. 
The true open system dynamics will be well 
described by equation (\ref{lindb})
as long as terms of the order
$\omega\tau$ and $\gamma\tau$ can be neglected, where $\tau$ is again 
the environmental memory time.

Our aim is a `post'-Markov master 
equation valid to {\it first} order in $\omega\tau$, $\gamma\tau$.
In what follows, we explain the main result and illustrate
it with several prominent physical problems, including 
a damped two-level system, the spin-boson model and the quantum Brownian 
motion model. We start with a
quantum system interacting with a bosonic oscillator environment 
with total Hamiltonian,
\beq
H_{\rm tot} = H
+ \sum_\lambda g_\lambda (L a_\lambda^\dagger + L^\dagger a_\lambda)
+ \sum_\lambda \omega_\lambda a_\lambda^\dagger a_\lambda.
\label{Htot}
\eeq
Here, $H$ is the Hamiltonian of the system and $L$ a system
operator describing the coupling to the environment.
The standard zero temperature bath correlation function of model
(\ref{Htot}) is
\beq\label{bathco}
\al(t,s)=\sum_{\la}g_{\la}^2e^{-i\om_{\la}(t-s)}.
\eeq
Its decay, as a function of the time delay $t-s$, defines the 
`memory' or correlation time $\tau$ of the environment.
We choose to normalize the coupling constants $g_\lambda$ in 
(\ref{Htot}) such that $\alpha(t,s)$ is normalized
($\int \alpha(t,s) ds = 1$) which means that the overall coupling 
strength is determined by a parameter hidden in
the coupling operator $L$ in (\ref{Htot}).

A lengthy derivation\cite{YDGS} based on a non-Markovian
quantum trajectory approach developed recently \cite{DGS}, shows
that to first order in the memory time $\tau$ the reduced density 
operator of model (\ref{Htot}) evolves according to
the first-order `post'-Markov master equation,
\beqa
\dot\rh_t =
&-&i[H,\rh_t]\label{mas1}\\
&+& g_0(t)[L, \rh_t L^\da] + g_0^\ast(t)[L\rho_t, L^\da]\nonumber \\
&+& ig_1(t)[L^\da,[H, L]\rh_t] -ig_1^\ast(t)[\rh_t[L^\da,H],L]\nonumber \\
&+& g_2(t)[L^\da, [L^\da,L]L\rh_t] +g_2^\ast(t)[\rh_t L^\da[L^\da,
L],L],\nonumber
\eeqa
where
\beqa
g_0(t)&=&\intt\al(t,s)ds\label{g0},\label{o1}\\
g_1(t)&=& \intt\al(t,s)(t-s)ds\label{g1},\label{o2}\\
g_2(t)&=& \intt\int^s_0\al(t,s)\al(s,u)(t-s)duds.\label{o3}          
\label{g2}
\eeqa
The master equation~(\ref{mas1}) is the central theme of this paper. 
Note first that according to our convention, the function $g_0(t)$ is 
of the order one. Therefore, as in (\ref{lindb}), the first two 
lines in (\ref{mas1})
define the time scales $\omega^{-1}$ ($H$) and $\gamma^{-1}$ 
($L^\dagger L$).
The two functions $g_1(t)$ and $g_2(t)$, however, are
of the order of the memory time $\tau$. Therefore,
the third and fourth line in (\ref{mas1}) are smaller than the
first and second line, respectively, by a factor 
$\gamma\tau$.
As $\tau\rightarrow 0$, the third and forth line 
in (\ref{mas1}) become negligible,
the real part of $g_0(t)$ tends to $\frac{1}{2}$, and
the contribution of the imaginary part of $g_0(t)$ can be absorbed by a
renormalized Hamiltonian: our `post'-Markov
master equation (\ref{mas1})
reduces to the Lindblad master equation (\ref{lindb}) for 
vanishing memory time $\tau$.
The third and fourth lines, which arise due to short but finite 
correlation time, represent the new first-order non-Markovian corrections.
The time-dependent coefficients $g_i(t)$
describe an important {\it initial slip} 
on the time scale of the memory time $\tau$ \cite{Haake}. Note that (\ref{mas1}),
assuming that the initial state is factorable,
is valid for zero temperature and also for finite temperature if $L = L^\dagger$.
General finite temperatue post-Markov master equation can be obtained direclty
from the first order perturbation of the finite temperature quantum state diffusion(QSD) 
equation\cite{DGS}.	


As a first example, we consider a damped two-level system:
$H=\frac{\om}{2}\si_z,\> \>L=\sqrt{\gamma}\si_-\;$. 
For the sake of simplicity,
we phenomenologically choose an exponentially decaying correlation 
function $\al(t,s)=\frac{1}{2\tau}\e^{-|t-s|/\tau}$ with
memory time $\tau$.
Since this model can be solved exactly\cite{AH,DGS}, we are able
to compare results from our post-Markov master equation
(\ref{mas1}),
\beqa
\dot\rh_t
&=&-i{\om\over 2}[\si_z,\rh_t]-i\om(\gamma g_1(t))
[\si_+\si_-, \rh_t]\nonumber\\
&& +\gamma(g_0(t)+\gamma g_2(t))(2\si_-\rh_t\si_+\nonumber\\
&& -\{\si_+\si_-,\rh_t\}),
\label{ze1}
\eeqa
with the exact result. It here happens that the post-Markov master
equation~(\ref{ze1}) is of Lindblad form~(\ref{lindb}) with
time-dependent coefficients
(the real $g_1(t)$ gives rise to a time-dependent 
frequency shift). Therefore we know that $\rho_t$ remains a
proper density operator for all times and all parameters.
As seen in the next two examples this property is not
a generic feature of the post-Markov master equation
(\ref{mas1}).
To illustrate the limits of the Markov approximation we compare
in Fig. 1 the average $\LJ\si_y\RJ$ obtained from the post-Markov 
master equation (dashed curve), from the
Markov master equation (dotted curve) and from the exact master 
equation (solid curve)
for the parameters $\omega=\gamma$ and $\gamma\tau=0.2$.
We see that the post-Markov 
master equation gives a much better result than the Markov master
equation as expected for a significant memory time of $\gamma\tau=0.2$.


So far, our results assume a zero temperature environment.
As stated before,  it turns out \cite{DGS}, that for a selfadjoint coupling 
operator $L=L^\dagger$ the same analysis holds, with
$\alpha(t,s)$ in (\ref{bathco}) replaced by the
finite temperature expression 
$\alpha(t,s) = \sum_\lambda g_\lambda^2
[\coth\left(\frac{\omega_\lambda}{2kT}\right)\cos\omega_\lambda(t-s)
-i\sin\omega_\lambda(t-s)]$ $(\hbar=1)$.
A finite temperature introduces an additional time scale 
$1/kT$ but we here
focus on the high-temperature limit $kT\gg \tau^{-1}$,
thus disregarding the new time scale. For
an Ohmic bath we obtain the standard correlation function
\begin{equation}\label{ohmalpha}
\alpha(t,s) = 2kT\Delta(t-s) + i \dot\Delta(t-s)
\end{equation}
where $\Delta(t)$ is a smeared out delta function, decaying on the
memory time scale $\tau$ (the inverse of the cutoff frequency), 
and $\dot\Delta$ is its
time derivative. The precise shape of $\Delta(t)$ 
depends on the type of high-frequency cutoff 
chosen in the sum (\ref{bathco}).
Due to the somewhat singular $\dot\Delta(t)$ term in (\ref{ohmalpha}),
the coefficient $g_1(t)$ in (\ref{o2})
is no longer of the order
of the memory time $\tau$ but turns out to be of the same order as
the zero order term $g_0$, and
both terms in (\ref{mas1})
are relevant in the $\tau\rightarrow 0$ limit.
The forth line in (\ref{mas1}) involving the function $g_2(t)$ 
vanishes for $L=L^\dagger$. Note that we have slightly
       changed our convention in order to be in line with standard 
       notation: the correlation function (9) is here normalized 
       such that the coefficient of the {\it imaginary} part, relevant 
       for the damping, is unity.

Let us consider the high-temperature spin-boson model \cite{LCDF}.
The system
Hamiltonian is
$H=-\frac{\omega}{2}\si_x + 
\frac{\Omega}{2}\si_z,$
where $\omega$ is the tunneling
matrix element and $\Omega$ depicts the bias of the system.
The coupling operator is $L=\sqrt{\gamma} \si_z$.
The `post'-Markov 
master equation can be obtained directly from (\ref{mas1}):
\beqa
\dot\rh_t=
&-&i[H,\rh_t]\nonumber\\
 &+&\gamma g_0(t)\si_z\rh_t\si_z-\gamma g_0(t)\rh_t + H.c.\nonumber \\
&-&i\omega\gamma g_1(t)\si_x\rh_t -\omega\gamma 
g_1(t)\si_y\rh_t\si_z +H.c.
\label{zm}
\eeqa
After an initial slip ($t\gg\tau$), the $g_i(t)$ approach
their asymptotic values and the density operator evolves
according to
\beqa
\dot\rh_t=
&-&i[H,\rh_t]
 +2kT\gamma\si_z\rh_t\si_z-
2kT\gamma\rh_t\nonumber \\
&+&\omega\gamma\{\si_x,\rh_t\} -i\omega\gamma\si_y\rh_t\si_z\nonumber\\
&+&i\omega\gamma\si_z\rh_t\si_y .
\label{zm1}
\eeqa
Obviously, both equation (\ref{zm}) and its asymptotic form (\ref{zm1})
are not of Lindblad form~(\ref{lindb}) due to the presence of 
the last three terms. 
Therefore, as a Markov equation with constant coefficients, 
we cannot
expect (\ref{zm1}) to preserve the positivity of $\rho_t$ if applied to
an arbitrary initial density operator. Our derivation
shows that for the asymptotic equation (\ref{zm1}) to be
meaningful, one has to use an {\it effective} initial
condition \cite{Haake} obtained by fully determining the 
initial slip arising from the time dependent coefficients $g_i(t)$
in the true master equation (\ref{zm}).

We show numerically that (\ref{zm})
preserves the positivity of the 
density matrix while the asymptotic equation (\ref{zm1}) applied
to the same initial condition fails to do so.
The positivity of $\rho_t$ is
equivalent 
to the condition $\vert\vert\langle\vec\sigma\rangle\vert\vert\leq 1$, 
where $\langle\vec\sigma\rangle={\rm Tr}(\vec\sigma\rh)$ is the
Bloch vector. In Fig. 2 we plot the norm of the
Bloch vector using the time-dependent master equation (\ref{zm}) (solid
curve) and using its asymptotic form (\ref{zm1}) right from
the start (dotted curve). 
Clearly, the latter loses 
positivity for some initial states on short time scales, whereas
the full post-Markov master equation (\ref{zm}) preserves it
for all times as numerically confirmed for a wide range of parameters
in the region $kT\gg\Lambda=\tau^{-1}$, where $\Lambda$ is the
cut-off frequency of the heat bath.
Note however that if the correlation time is too large, then the post-Markov
master equation can also produce non-positive results as should be expected
from its derivation.



Note that this simple model is the 2-level analog of the 
quantum Brownian motion model where we have
$H=\frac{1}{2}p^2+ V(q)$ and $L=\sqrt{\gamma} q$.
The post-Markov master equation (\ref{mas1}) for this case
reads
\beqa
\dot\rh_t =
&-&i[H,\rh_t] - \gamma g_{0R}(t)[q,[q,\rh_t]]
-i\gamma g_{0I}(t)[q^2,\rh_t]\nonumber\\
 &+&\gamma g_{1R}(t)[q,[p,\rh_t]]+i\gamma g_{1I}(t)[q,\{p,\rh_t\}]\label{fi}
\eeqa
where the coefficients $g_{iR}(t), g_{iI}(t)$ are the real and
imaginary parts of $g_{i}(t),(i=0,1)$, respectively.
In the special 
case when $V(q)$ is a quadratic potential, it is reassuring that our 
master equation (\ref{fi}) coincides with the first order 
expansion of the exact Hu-Paz-Zhang master equation \cite{HPZ}. 

As in the previous example,
for the high temperature Ohmic correlation function (\ref{ohmalpha}),
the functions $g_i$ approach their asymptotic values
after an initial slip on the time scale $\tau$ 
and (\ref{fi}) becomes
the standard quantum Brownian motion 
(Caldeira-Leggett~\cite{CaLe}) master equation
\beq
\dot\rh_t = 
-i[H^\prime, \rh_t] -i\frac{\gamma}{2}[q, \{p,\rh_t \}] 
-\gamma kT[q, [q,\rh_t]],
\label{cale}
\eeq 
where $H^\prime$ is a cutoff-dependent renormalized Hamiltonian
and where we dropped a term involving $\tau p$ with respect to 
a similar term proportional to $q$. 

It is known that the non-Lindblad master equation (\ref{cale}) may 
violate the positivity of the density operator. 
Our full post-Markov master equation (\ref{fi}) is also a 
non-Lindblad equation but with time-dependent coefficients. As in 
the case of the spin-boson model, their 
time dependence can assure the preservation of the state's positivity
for a wide range of parameters in the validity region of Eq. (\ref{zm}).
In (\ref{fi}), the coefficient $g_{1I}(t)$ of the
dissipative term is zero at $t=0$ and its time derivative vanishes, too. 
The diffusion coefficient $g_{0R}(t)$ also vanishes
but its initial derivative is positive. Thus the initial phase of the
evolution is dominated by diffusion. This mechanism,
as is well known in the exact model of Ref. \cite{HPZ}, may guarantee
the positivity of the density matrix for short times as well as at later
times when the dissipation enters. In contrast, the 
asymptotic ($t\gg\tau$) Caldeira-Leggett
master equation (\ref{cale}), if used right from the start ($t\ge 0$),
will immediately violate the positivity of a distinguished class of 
initial density matrices due to the constant dissipative term.

It is interesting to stress the close  connection between the post-Markov
master equation (\ref{mas1}) and the QSD
equation corresponding to (\ref{Htot}) (see \cite{YDGS}). Indeed,
the former was derived from the first order perturbation of the
latter. Consequently, the post-Markov master equation (\ref{mas1}) can be 
simulated by using the powerfull quantum trajectory techniques provided 
by QSD\cite{YDGS}.

 To sum up, we have presented a post-Markov master equation for
 the evolution of open quantum systems. We have illustrated
 its great potential using several examples and we have further shown
 numerically that, for a wide range of parameters, the post-Markov master 
 equation preserves the positivity of the density operator 
 while its asymptotic version, disregarding the initial slip,
 fails to do so. This opens the road to applications of the 
 post-Markov master equation (\ref{mas1}) to phenomena, such as
 quantum decoherence and dissipation, photonic bandgaps, atom
 laser and more generally atom-field interactions
 in the weakly non-Markovian regimes.



 We thank Ian C. Percival for useful suggestions and Bernard 
 Gisin for help in numerical simulations. TY and NG would 
 like to thank for support from the Swiss National Science Foundation and the 
European TMR Network ``The physics of quantum
information" (via the Swiss OFES). WTS thanks the Deutsche 
Forschungsgemeinschaft for support through the Sonderforschungsbereich 237
``Unordnung und gro{\ss}e  Fluktuationen''.

\section*{Captions of Figures}

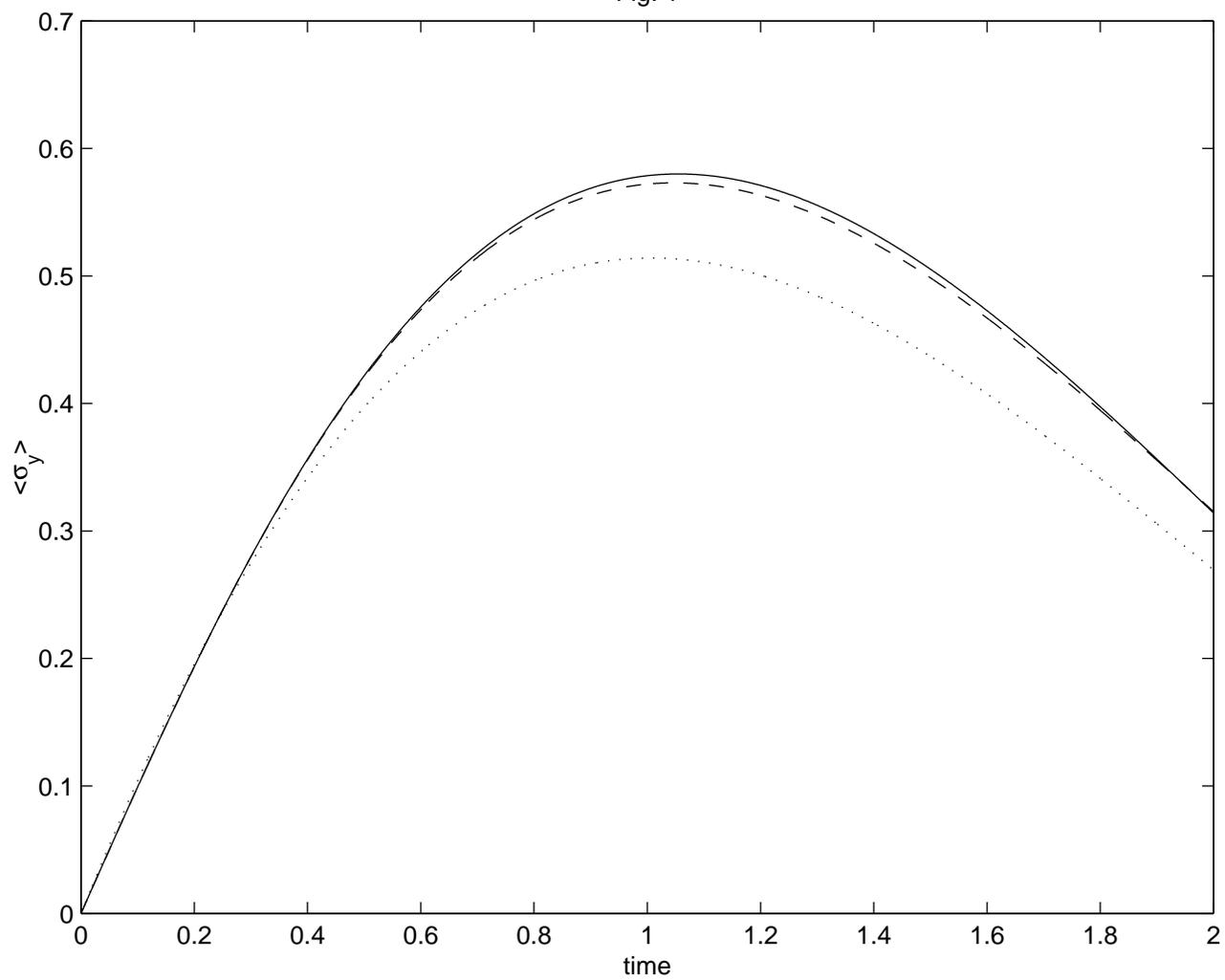
FIG. 1.
Illustration of the limit of the Markov approximation 
(dotted curve) for a damped two-level system.
The dashed curve is our post-Markov result, the 
solid curve is the exact result. The initial state
is $|\psi_0\RJ=|-\RJ$.
The parameters are  $\omega=\gamma, \gamma\tau=0.2$.

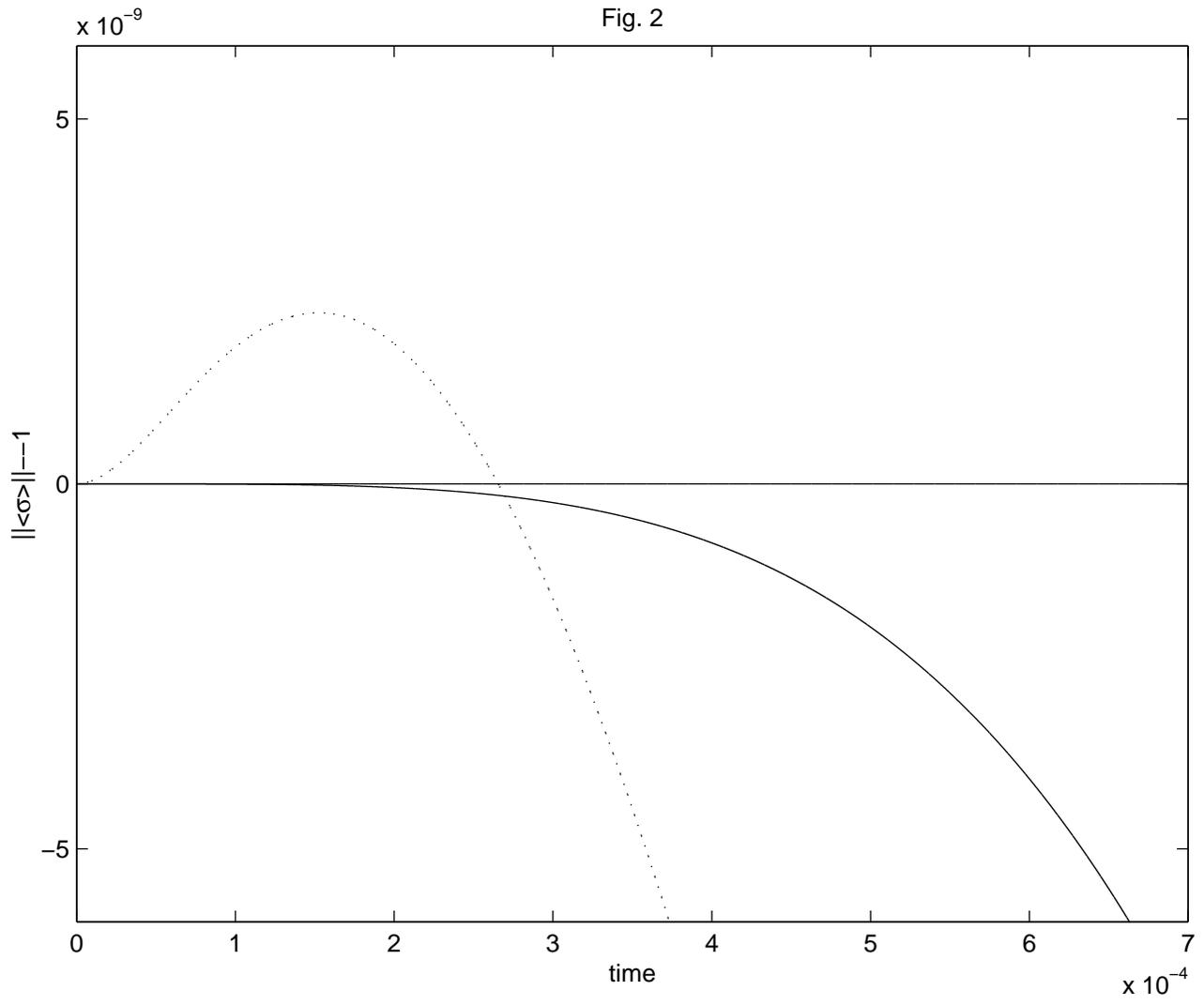
FIG. 2. 
Norm of the Bloch vector for the spin-boson model evolved with
the post-Markov equation (\ref{zm}) (solid curve) and evolved with
the asymptotic equation (\ref{zm1}) (dotted curve).
As the latter leads to a norm greater than one, it violates
the positivity of $\rho_t$.
The initial state is $|\psi_0\RJ=|+\RJ$ and 
the parameters are $\Omega=0, \omega\tau=0.01, kT\tau=20, 
\gamma=0.3\omega$.


\end{document}